# Half-Saturation Constants in Functional Responses

Christian Mulder[1,*] and A. Jan Hendriks[2]

[1] National Institute for Public Health and the Environment, Box 1, 3720BA Bilthoven, The Netherlands

[2] Department of Environmental Science, Radboud University Nijmegen, Box 9010, 6500GL, Nijmegen, The Netherlands

**arXiv.org > q-bio > q-bio.QM**

**ABSTRACT** – Intake of elemental nutrients by plants and food by animals is often considered to be a hyperbolic or sigmoid function of the resource. In these relationships, the half-saturation constant $K_m$ [kg·km$^{-2}$, kg·l$^{-1}$], i.e. the resource availability at which half of the maximum intake is reached, determines the outcome of models and contributes to explain the life-strategies of species. As data on this parameter are rather scarce, our investigation aims (1) to provide an overview of the half-saturation constants reported in current literature and (2) to explore the consistency of the collected data with body size. First, a meta-analysis was conducted on reviews and original studies published in worldwide literature. In total, 338 half-saturation constants were collected from bacteria to ungulates. Most studies focused on algae and invertebrates, whereas some included fish, birds and mammals. Next, the pooled half-saturation constants obtained were linked to body size, using ordinary linear regressions.
**Contact:** christian.mulder@rivm.nl, A.J.Hendriks@science.ru.nl.

## 1   ABSORPTION AND CONSUMPTION

Absorption of elemental nutrients by plants and ingestion of food by animals are two crucial processes in the understanding of ecosystem functioning, including the assessment of the effects of anthropogenic interference such as overgrazing, overfishing and eutrophication. Generally, the rate of intake is considered to increase with nutrient concentration and food density, until it levels off due to some kind of saturation. This relationship has been described by a large number of mathematical equations. For instance, over 40 different functions have been proposed for consumption [1,2]. By contrast, empirical support for these relationships is limited to few taxonomic groups. Even more, the lack of data is unlikely to be reduced substantially by additional species-specific observations because of financial, practical and ethical restrictions.

Ecological assessments that aim to cover a broad taxonomic diversity often contain intake functions with parameters for which only some values are available. Fortunately, most absorption and ingestion experiments have been examined using a single function that relates the intake rate constant k to the nutrient concentration and food density in the environment N according to Equation 1:

$$k = \max(k) \cdot \frac{N^{\beta}}{N^{\beta} + K_m^{\beta}}$$

The aim of the present study was to provide an overview of half-saturation constants for nutrient absorption and food ingestion ($K_m$) and explore possible relationships to the trophic level, the taxonomic group and the size of species. To achieve this, data were related to species' body mass $m$ with $\log_{10}$-$\log_{10}$ linear regressions derived from Equation 2: $K_m = \gamma \cdot m^{\kappa}$

If the resources are scarce, i.e. N < $K_m$, absorption and ingestion k rate constants [kg · kg$^{-1}$ · d$^{-1}$] linearly increase with respectively the concentration of the elemental nutrients and the density of the food resource N [kg · km$^{-2}$, kg · l$^{-1}$] (Figure 1). If resources are (more) abundant, i.e. N > $K_m$, intake k levels off to the maximum value max(k) due to transport and transformation delays such as nutrient allocation in plants or food digestion in animals. The exponent β indicates either inhibition (β < 1), yielding a hyperbolic curve, or facilitation (β > 1), yielding a sigmoid curve. In biochemistry, the relationship describes the transport and transformation of substances, such as oxygen or glucose, either without (β ≤ 1) or with (β > 1) allosteric effects [3,4]. In microbiology and plant sciences, the hyperbolic equation is used for nutrient intake [5]. Ingestion of food by animals is described by either a Type II (β = 1) or Type III (β = 2) functional response, also reflecting independence and facilitation, e.g. due to experience in search or handling [6]. Equation 1 is also used to forecast the uptake of toxicants [7].

## 2   HALF-SATURATION

The half-saturation constant $K_m$ [kg · km$^{-2}$, kg · l$^{-1}$] represents the concentration or density at which half of the maximum intake rate [½ · max(k)] is reached, independently of β. Low $K_m$ values apply to plants and animals that acquire resources rapidly at low concentrations and densities, high values are noted for inefficient organisms. For instance, $K_m$ values for nutrient absorption by phytoplankton increase along a gradient from oligotrophic oceans to eutrophic estuaries, even within the same species [8]. It suggests that the occurrence and adaptation of organisms to the level (and, hence, the stoichiometric quality) of resources in their environment is reflected in the value of the half-saturation constant $K_m$.

---

[*] To whom correspondence should be addressed.

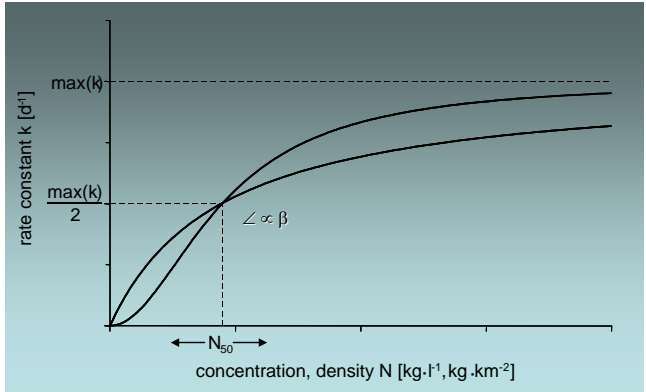

**Figure 1.** Absorption of elemental resources and ingestion of food k [$d^{-1}$] versus nutrient concentration and food density N [$kg \cdot l^{-1}$ or $kg \cdot km^{-2}$], respectively, according to Equation 1 with a hyperbolic ($\beta < 1$) and sigmoid ($\beta > 1$) response set by the half-saturation concentration or density $N_{50}$ [$kg \cdot l^{-1}$ or $kg \cdot km^{-2}$] at which half of the maximum rate max(k) is reached.

Few experiments have been carried out to obtain $K_m$ directly. Instead, the half-saturation constant is often derived indirectly by fitting the output of ecological models as a whole to field data on population dynamics. This pragmatic approach is adequate in cases where uncertainty in the intake function is known to dominate the variability of the output. In other cases, parameter values have to be derived from enrichment experiments and feeding trials, independently of field dynamics. Moreover, parameter calibration with laboratory experiments and model validation with field surveys is to be preferred for good modelling practice. The selection of appropriate values might be improved by relating the half-saturation constants for species to species' properties, such as trophic level, and to species' traits, such as average body size. Of the two coefficients in Equation 1, the allometric regressions for maximum rates of nutrient absorption and food ingestion max(k) have been obtained, covering different taxonomic groups [9–17]. Besides for zooplankton [14], other body-mass regressions for $K_m$ have not been reported yet. In contrast to a comparable meta-analysis on Types I and II functional responses [18], where more attention was provided to vertebrates, we decided here to focus more on plants and invertebrates (compare 27 % of the entries for vertebrates in [18] vs. 12 % of the entries for vertebrates in the present paper).

**Plants**

On average, half-saturation constants $K_m$ [$kg \cdot l^{-1}$] for plant uptake of Nitrogen and Phosphorus did not significantly deviate from each other (Table 1). For both elements the values were mostly between $10^{-9}$ to $10^{-7}$ $kg \cdot l^{-1}$ (Figure 2). For individual studies [19–22], half-saturation constants for different species were generally within one order of magnitude. $K_m$ for autotrophs scaled weakly (Table 1) but significantly to size ($p < 0.05$), although the nutrient absorption by plants becomes size-independent as soon we focus strictly on microphytes ($0.3 < p < 0.8$). The allometric relationship was further strengthened by the low value for Phosphorus uptake by bacterial cells ($m \approx 10^{-15}$ kg, [23]). Levels for macro-algae, herbs and forbs, and tree seedlings were in the same range of magnitude, with the exception of the low Phosphorus concentration of $2 \times 10^{-9}$ $kg \cdot l^{-1}$ noted in *Laminaria japonica* kelp forests [24].

**Cold-blooded animals**

Half-saturation constants $K_m$ for ingestion of algae by different invertebrate species groups were largely within the range of $10^{-7}$ to $10^{-5}$ $kg \cdot l^{-1}$ (Figure 3). The averages for ciliates and mollusc larvae were lower than other. Values for different species tested under the same experimental conditions roughly varied one order of magnitude (extensively reviewed in [14]). As an indication of the intraspecific variability, values for equally-sized *Daphnia magna* were within the same order of magnitude (encircled '(a)' in Figure 3). The food density at half of the maximum ingestion $K_m$ did not increase with size if all invertebrate herbivores were included into the regression (Table 1). Per species group, however, $K_m$ scaled to mass *m* with exponents in the range of 0.26–0.70. While trends for these groups were present ($0.03 < p < 0.20$ vs. $0.09 < p < 0.64$), size scaling was not detectable in ciliates ($p = 0.7$, $r^2 = 0.01$).

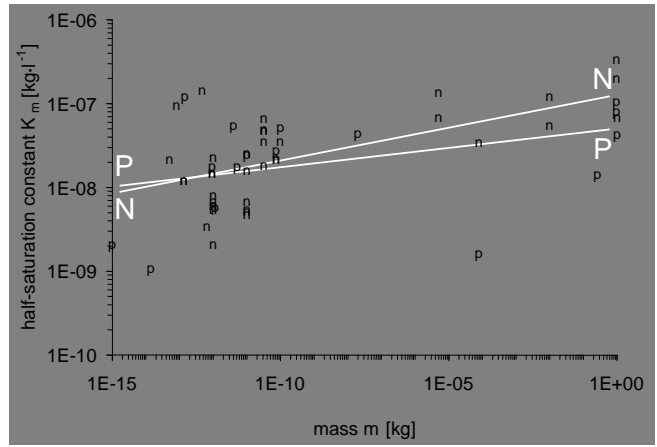

**Figure 2.** Half-saturation constants $K_m$ [$kg \cdot l^{-1}$] vs. plant mass *m* [kg] for nutrient absorption. Data points (lower case letters) and linear regressions (upper case letters) thereof with N,n = Nitrogen, P,p = Phosphorus.

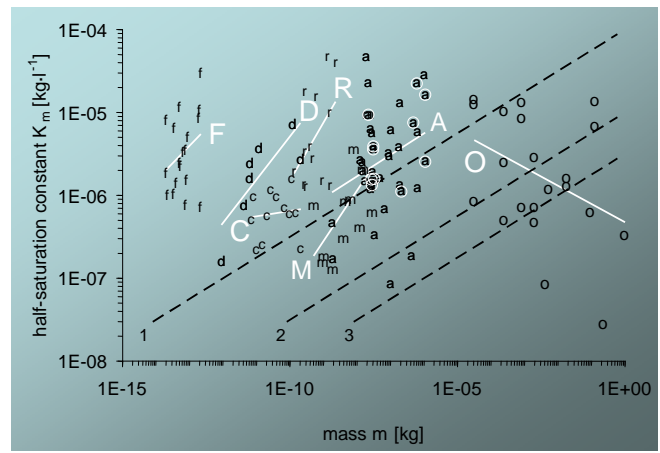

**Figure 3.** Half-saturation constants $K_m$ [$kg \cdot l^{-1}$] vs. animal mass *m* [kg] for ingestion of food by aquatic herbivorous invertebrates and piscivorous fishes. Data (lower case letters) and regressions (solid lines with upper case letters) for **F**,f = Sarcomastigophora, **D**,d = Dinoflagellata, **C**,c = Ciliophora, = **R**,r Rotifera, **M**,m = Mollusca, **A**,a = Arthropoda, **O**,o = Osteichthyes. In addition, the maximal food densities [$kg \cdot l^{-1}$] as represented by three trophic levels [**1** (algae), **2** (detritivores-herbivores) and **3** (carnivores)] from [63] were calculated assuming $1\ kg \cdot l^{-1} \sim 10^{10}\ kg \cdot km^{-2}$ and consumer-to-resource body-mass ratios ($m_i/m_{i-1}$) of $10^4$ for planktivores (phyto- and zooplanktivores) and $10^2$ for piscivores (dashed lines: **1, 2, 3**).

Half-saturation constants for tilapia were not plotted because this fish feeds on blue-green algae (Cyanobacteria). Still, values of 2× up to 6×10$^{-5}$ kg · l$^{-1}$ are in the upper edge of the range noted for herbivorous invertebrates [25]. Experiments with piscivorous fishes were carried out with immature individuals: resulting values are mostly in the 10$^{-7}$ – 10$^{-5}$ kg · l$^{-1}$ range, besides extremely low values for herring [26] and lake trout [27]. There was no consistent difference between prey trials with zooplankton and fish as resources. In contrast to other allometric regressions derived in the present study, half-saturation for fish decreased weakly with mass.

**Warm-blooded animals**

Data sets for warm-blooded herbivores included mammals ranging from lemming to bison, while the carnivorous group consisted of several bird and mammals species (Figure 4). Regression analysis indicated size dependence for both trophic levels (0.02 < p < 0.20, 0.31 < r$^2$ < 0.41). Average levels for herbivorous homeotherms were significantly higher than those for carnivorous equivalents (Table 1). For both, variability between the species observed in the same study was less than one order of magnitude [28,29].

Values for herbivores applied to different species, with exception of sheep (at 53 kg in Figure 4). The half-saturation constants reported in these studies varied a factor of 1.8 [28,30]. All data on (semi)-arid terrestrial consumers were below the regression line [28,30,31]. Data from experiments with artificially controlled plots were not included in the analysis because observations were between about 55 to 22,000 kg · km$^{-2}$, clearly below field levels [32]. Values for browsers were excluded because functional responses were related to bite size rather than to plant density [1].

The low value for carnivores applied to the Arctic Fox, being *Alopex lagopus* with its 5 kg much more efficient than other Arctic predators [29]. The half-saturation constant for lizards feeding on grasshoppers was not plotted because Figure 4 shows only warm-blooded species. Still, the value of 200 kg · km$^{-2}$ noted for reptiles was in the range for homeotherms feeding on vertebrates [33].

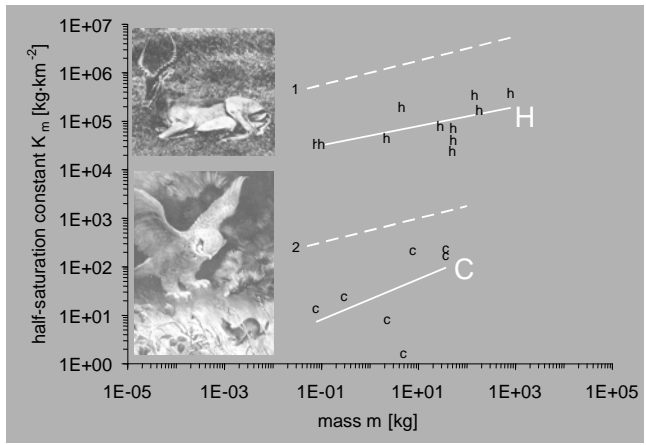

**Figure 4.** Half-saturation constants $K_m$ [kg · km$^{-2}$] vs. animal mass $m$ [kg] for ingestion of food by warm-blooded animals (Mammalia + Aves). Data (lower case letters) and regressions (solid lines with upper case letters) with **H**,h = herbivorous mammals and **C**,c = carnivorous mammals + predatory birds. In addition, maximal food densities [kg·km$^{-2}$] as represented by trophic levels **1** (vegetation) and **2** (herbivores) were calculated assuming consumer-to-resource body-mass ratios $m_i/m_{i-1}$ of 10$^4$ for herbivores and 10$^1$=10 for carnivores (dashed lines) as in donor-controlled ecosystems.

## 3 REGRESSION SLOPES

Although data for many species groups are scarce, some general patterns emerge from the analysis. All but two whole-taxon regressions showed that half-saturation constants increase with size, using significance levels that are common in allometric correlations (0.0001 < p < 0.20). Values for ciliates were independent of size, possibly due to the small body-mass range covered (p = 0.71). Fish data did not follow the pattern noted for other species. The deviation persisted even if differences between methods were excluded, because all data with $m < 2\times10^{-3}$ kg were obtained from the same experiment [34] (Figure 3), where attack rate and handling time scaled to predator's length L with exponents of 4.4 and -3.4. Their product was inversely related to the half-saturation constant, yielding $K_m \propto L^{-(4.4-3.4)} = L^{-1} \propto m^{-\frac{1}{3}}$, not so far from the $m^{-0.22}$ of Table 1. In these experiments, the diet breadth for both small and large fishes was held constant: this might explain the decrease of handling time with predator's size. Under field conditions, bigger fishes tend to select larger food items, likely to require more time for handling and corresponding to higher saturation constants.

Previous attempts to relate $K_m$ to size are rare. A negative slope was observed for Phosphorus intake by algae [15]; however, the same study reported an exponent of 0.30 for the maximum intake rate max($k_n$) while the ratio max($k_n$)/$K_m$ scaled to 0.32, suggesting an exponent of -0.30 / -0.32 = 0.02 instead of -0.28. Half-saturation constants $K_m$ for consumption of algae by rotifers scaled with a slope of 1.22. A previous analysis on phytoplanktivores yielded an exponent of -0.03 [14], close to the aforementioned exponent of 0.02. The present analysis was largely based on the same data for invertebrates and confirmed the independency of size. However, this analysis showed that $K_m$ seems to increase with consumer's size as soon as smaller phylogenetical groups are considered.

How can we explain an increase of $K_m$ with the faunal body size? To that end, we might compare the slopes noted in general for $K_m$ [kg · km$^{-2}$] to those noted for food density N [kg · km$^{-2}$]. Following the energy equivalency rule [35–37], mass values N of all species occurring within one trophic level are expected to scale approximately inverse to rate constants, i.e., $N \propto m^{1/4}$ [35–38]. The slopes of the regressions for half-saturation constants of invertebrates feeding on algae varied between 0.07 and 0.70 (Table 1), whereas average and maximal phytoplankton density scaled to algal size with 0.22 to 0.47 [39–42].

Vegetation density at half the maximum grazing rate increased with size to the power of 0.20 (Table 1) where values between 0.21 and 0.33 were noted for plant density [43–46]. Densities of mammalian herbivores themselves scaled to size with an exponent between 0.17 and 0.44 [35,36,47–50]. The slope noted for the density of herbivores at half the maximum grazing rate was 0.57, rather out of this range (Table 1). However, $K_m$ values generally increase with faunal body size in the same way as the density of their prey species, suggesting that the functional response of animals is adapted to the food that they encounter. Both the half-saturation constant and the density are relatively low for small animals and high for large animals.

The weaker scaling of half-saturation constants for absorption of elemental nutrients by plants cannot always be attributed to a comparable mechanism, because due to mutualism, vascular plants do not necessarily imply "more" macronutrients than algae. However, theoretical models confirm the empirical observation that half-saturation constants for nutrients increase at least with algal size, due to various processes like dilution [51]. Indirect support comes from the notion that minimum nutrient requirements increase with algal size too, reflecting a similar rise in dry matter content [52].

## 4   AVERAGES AND INTERCEPTS

The half-saturation constants for absorption [kg Nitrogen · l$^{-1}$ and kg Phosphorus · l$^{-1}$] were about two orders of magnitude smaller if compared to those for ingestion [kg Food · l$^{-1}$]. If expressed on a nutrient basis, half-saturation constants are at the same level. Nitrogen and Phosphorus contents in typical biota seem to be 2 and 0.3 %, respectively, so that elemental levels are expected to be 50 and 330 lower than food densities. Per trophic level, $K_m$ values for nutrients and food uptake generally varied within two orders of magnitude (Figures 2–4). This variability possibly reflects divergent strategies in resource acquisition or different conditions between studies. Data on *D. magna* indicated that differences between conditions contribute by a factor of 10 to the observed variability. Differences between species examined within the same study account for another order of magnitude.

Intercepts for the average density [kg · km$^{-2}$] of a species are variable, depending on availability of sunlight, water, nutrients, and to a lesser extent to biodiversity. In speciose ecosystems under optimal conditions, the total number of all species occurring in the same trophic level [*sensu* 53] can be used as a reference for the half-saturation. This upper level reflects the total biomass within one trophic level as a function of the species' body-size average or the total biomass of the population of dominant species. The typical values noted in a meta-analysis of empirical regressions were converted and plotted as a function of consumers' body mass. We may note in Figure 3 that the total phytoplankton density as a function of herbivore's size was at the same level as the $K_m$ values for ciliates, mollusc larvae and arthropods (dashed line 1 versus solid lines C, M, A). Levels for flagellates and rotifers were substantially higher (solid lines F, D, R). Similar discrepancies are calling for species-specific analyses on consumer-resource body-mass ratios, nutritional quality and other elemental factors [14,18,23,54–56].

Maximal vegetation density was at half the grazing rate one order of magnitude higher than grass density in Figure 4 (dashed line 2 versus solid line H). A similar difference was noted for smaller herbivores caught by predators (dashed line 3 versus solid line C). If the maximum density may be considered to reflect the carrying capacity K for the trophic level concerned, $K_m/K$ ratios can be calculated to be approximately 1/10 for terrestrial homeotherms. Independent data on mammals indicate ratios around 1/10 [56–59]. The value of $K_m$ itself, as well as its value relative to the carrying capacity $K_m/K$ is crucial for the consumer–resource dynamics.

## 5   IMPLICATIONS

Many ecological models that are applied to explore options or support decisions in management contain parameters that have not been determined empirically for most taxonomic groups. Often, these parameters are obtained by varying them simultaneously until the discrepancy between the predicted and observed population dynamics is minimal. Unfortunately, calibrating parameters by comparing model and field dynamics only gives indirect values, possibly influenced by other factors in the model. The credibility of these models can be increased by deriving parameters values from independent observations of the underlying processes. As an alternative, one may link parameters to well-known suites of traits and properties of species and communities.

The present analysis focused on the half-saturation constant $K_m$, an important parameter in ecological and environmental models on sustainable fisheries, rangeland management, pollution, and so on. To a certain extent, $K_m$ can be estimated as a function of size with an almost "physically universal" exponent of, e.g., ¼ and an intercept that is determined by the chosen $K_m/K$ ratio, e.g. 1 for aquatic invertebrates and 10 for homeotherms [e.g. 60–62]. Alternatively, if field calibration is preferred, the present analysis may help to underpin and understand the value derived for the half-saturation constant $K_m$. In a comparable way with growth rates and maturity age that scale to size with exponents that are opposite to each other (-¼ vs. ¼), $K_m$ values need to be coherent with related parameters.

We can now put measurements of the half-saturation constant $K_m$ in a broader framework. These values are often explained in terms of physiological characteristics of the consumers, including, e.g., handling and digestion time [2,18], but while these processes are undoubtedly important at the level of individuals, this analysis suggests to a certain degree a kind of macroecological nature.

Species size explained between 1 and 64 % of the variability of $K_m$. Obviously, this fraction can be increased by relating half-saturation to supplementary characteristics of the species and communities concerned. While size remains an important trait, other traits may be also important. Size is not the dominant factor predicting the half-saturation. Future investigations may shed light on trait-derived factors, if more empirical data become available, especially on taxa with deviating trends. Nevertheless, the present study already demonstrated that resource and consumer body masses are extremely important in understanding variability in the functional response for various types of resources and species.

Christian Mulder and A. Jan Hendriks

**Table 1** - Half-saturation constants for absorption of nutrients by plants and ingestion of food by animals, categorized according to resource and consumer. The number of data (n), the geometric average with its confidence interval (µ, 95% C.I.), the allometric regression $K_m = \gamma \cdot m^\kappa$ and statistical significance were provided for each subset. Values referring to the same species, size or resource were averaged geometrically if taken from equivalents. Nutrients were reported as the nitrate, ammonium, nitrogen and phosphorus concentrations in water. Food densities at half the maximum intake rate $K_m$ for aquatic species were mostly measured in a volume of water [kg·l$^{-1}$]. To obtain a consistent data set, aqueous values were discarded if expressed per unit of area. Half-saturation constants $K_m$ for terrestrial animals usually referred to the numerical abundance [km$^{-2}$] or the wet biomass [kg·km$^{-2}$] of food in a region. In the discussion, half-saturation constants for food were converted to allow aquatic [kg·l$^{-1}$] and terrestrial [kg·km$^{-2}$] species to be compared to each other and to other density parameters. Area units were converted to volumetric equivalents, assuming a water depth of 10 m, close to the geometric average of 5 and 100 m, noted in freshwater and marine systems. Areal density of phytoplankton communities scaled to size according to $10^7 \cdot m^{¼}$ [kg·km$^{-2}$], which corresponds to a volumetric concentration of $10^7/10^{10} \cdot m^{¼}$ [kg·l$^{-1}$]. To allow a direct comparison of nutrient concentrations and food densities, all organisms were assumed to have a molecular composition of $C_{106}H_{180}O_{46}N_{16}P$. As a result, 1 kg of wet biomass corresponds to 20%·16·14/2443 kg = 0.02 kg N and 20%·1·31/2443 kg = 0.0025 kg P dry matter. The $K_m$ values were then linked to body size, using ordinary regression analysis.

| resource | consumer | n | µ (95%-CI) | $K_m = \gamma \cdot m_i^\kappa$ | $r^2$ | p |
|---|---|---|---|---|---|---|
| nitrogen [kg·l$^{-1}$] | all plants | 44 | 3.0·10$^{-8}$ (2.1·10$^{-8}$–4.4·10$^{-8}$) | 1.2·10$^{-7}$·$m^{0.08}$ | 0.49 | <0.0001 |
| | bacteria and microalgae | 26 | 1.5·10$^{-8}$ (9.9·10$^{-9}$–2.3·10$^{-8}$) | 1.8·10$^{-7}$·$m^{0.09}$ | 0.04 | 0.34 |
| phosphorus [kg·l$^{-1}$] | all plants | 22 | 2.5·10$^{-8}$ (1.4·10$^{-8}$–4.5·10$^{-8}$) | 5.0·10$^{-8}$·$m^{0.05}$ | 0.18 | 0.05 |
| | bacteria and microalgae | 9 | 3.1·10$^{-8}$ (1.6·10$^{-8}$–6.0·10$^{-8}$) | 6.5·10$^{-8}$·$m^{0.03}$ | 0.01 | 0.77 |
| algae [kg·l$^{-1}$] | Sarcomastigophora | 18 | 3.3·10$^{-6}$ (1.9·10$^{-6}$–5.5·10$^{-6}$) | 1.1·10$^{0}$·$m^{0.42}$ | 0.10 | 0.21 |
| | Dinoflagellata | 7 | 1.7·10$^{-6}$ (5.4·10$^{-7}$–5.4·10$^{-6}$) | 8.4·10$^{-1}$·$m^{0.52}$ | 0.64 | 0.03 |
| | Ciliophora | 12 | 6.1·10$^{-7}$ (4.0·10$^{-7}$–9.1·10$^{-7}$) | 2.9·10$^{-6}$·$m^{0.07}$ | 0.01 | 0.71 |
| | Rotifera | 14 | 5.2·10$^{-6}$ (2.3·10$^{-6}$–1.2·10$^{-5}$) | 1.5·10$^{1}$·$m^{0.70}$ | 0.22 | 0.11 |
| | Mollusca | 12 | 7.1·10$^{-7}$ (3.1·10$^{-7}$–1.6·10$^{-6}$) | 8.5·10$^{-2}$·$m^{0.61}$ | 0.38 | 0.03 |
| | Arthropoda | 37 | 2.7·10$^{-6}$ (1.7·10$^{-6}$–4.4·10$^{-6}$) | 2.2·10$^{-4}$·$m^{0.26}$ | 0.09 | 0.08 |
| | all invertebrates | 99 | 2.1·10$^{-6}$ (1.6·10$^{-6}$–2.8·10$^{-6}$) | 3.8·10$^{-6}$·$m^{0.03}$ | 0.01 | 0.47 |
| animals [kg·l$^{-1}$] | Osteichthyes | 21 | 1.7·10$^{-6}$ (7.8·10$^{-7}$–3.8·10$^{-6}$) | 4.7·10$^{-7}$·$m^{-0.22}$ | 0.15 | 0.09 |
| grasses [kg·km$^{-2}$] | grazing Mammalia | 11 | 8.6·10$^{4}$ (4.5·10$^{4}$–1.7·10$^{5}$) | 4.9·10$^{4}$·$m^{0.20}$ | 0.41 | 0.04 |
| mammals [kg·km$^{-2}$] | pred Aves + carn Mammalia | 7 | 3.4·10$^{1}$ (5.8·10$^{0}$–2.0·10$^{2}$) | 1.5·10$^{1}$·$m^{0.57}$ | 0.31 | 0.20 |

Main references for: Protista, Mollusca, and Arthropoda [14]; Schizo-Phycophyta [19-22,24]; Spermatophyta [61-63]; Osteichthyes [26,27,34,64-66]; Mammalia [28-31,33,57,58,67-70].